\input epsf
\documentstyle[seceq,supplement]{ptptex}

\title{
Non-locality as an essential feature of brane worlds
}

\author{
Shinji {\sc Mukohyama}~\footnote{E-mail: mukoyama@schwinger.harvard.edu}
}

\inst{
Department of Physics, Harvard University\\
Cambridge, MA, 02138, USA
}

\recdate{
}

\abst{
Weak gravity in the brane world scenario with a scalar field in the bulk
between two end-of-the-world branes is investigated by using the
doubly gauge invariant formalism. We review the result that Einstein
gravity is restored in the low energy limit and that high-energy
corrections to Einstein gravity can be mimicked by higher derivative
terms. After that, it is argued that non-locality is one of essential
features of brane worlds. 
}

\begin{document}

\maketitle


\section{Introduction}


The idea that our four-dimensional world may be a timelike surface, or
a world-volume of a $3$-brane, in a higher dimensional spacetime has
been attracting a great deal of physical interests. As shown by
Randall and Sundrum~\cite{RS2}, the $4$-dimensional Newton's law of
gravity can be reproduced on a $4$-dimensional timelike hypersurface
with positive tension in a $5$-dimensional AdS background despite the
existence of the infinite fifth dimension.


However, if we consider another brane parallel to the first brane at 
finite distance then the weak gravity at low energy on either brane
becomes Brans-Dicke type~\cite{Garriga-Tanaka}. What is interesting is
that Einstein gravity is recovered if we further include a scalar field 
between the two
branes~\cite{Tanaka-Montes,Mukohyama-Kofman}. From the $4$-dimensional
point of view, these facts can be understood as the emergence of a
massless mode called radion in the former case and the stabilization of
the radion in the latter case. Namely, if we have two branes but no bulk
scalar then the inter-brane distance is arbitrary and, hence, the
corresponding mode should be massless. The massless mode plays
the role of the Brans-Dicke scalar field. If we include a bulk scalar
field which couples to the branes then the inter-brane distance is no
longer arbitrary and, hence, the would-be massless mode should obtain a
nonzero mass. Thus, the would-be Brans-Dicke scalar will disappear from 
the low energy description of brane gravity and, thus, Einstein gravity
is recovered.

The above understanding of the emergence of Brans-Dicke theory and the
recovery of Einstein gravity is appealing to our intuition based on
$4$-dimensional gravitational theory: our intuition is that low energy
gravity should always look like either Einstein or Brans-Dicke gravity 
and that all we have to do should be to investigate whether there
exists a massless scalar or not. Although this is true in the above
cases without or with just one bulk scalar field, we may not trust this
strategy too seriously for more general cases since we are dealing with
a $5$-dimensional system and it is rather non-trivial whether a
$4$-dimensional description is valid or not. Actually, in order to claim
that the weak gravity at low energy is Einstein, we have to show that
the effective $4$-dimensional gravitational coupling constants
$G_{N}^{(S)}$ and $G_{N}^{(T)}$ for scalar and tensor modes,
respectively, are the same~\cite{Mukohyama-Kofman}, where
these coupling constants are defined by 
%
\begin{eqnarray}
 \partial^{\rho}\partial_{\rho}h & = & 
  -16\pi G_{N}^{(S)}\tau, \nonumber\\
 \partial^{\rho}\partial_{\rho}h_{\mu\nu}^{TT} & = & 
  -16\pi G_{N}^{(T)}\tau_{\mu\nu}^{TT}. \label{eqn:GN}
\end{eqnarray}
Here, $h$ and $h_{\mu\nu}^{TT}$ are the trace and the transverse
traceless part of metric perturbation on the brane, and $\tau$ and
$\tau_{\mu\nu}^{TT}$ are those of stress energy perturbation of matter
on the brane. (In the main part of this paper, we shall deal with 
gauge-invariant quantities only and, thus, there is no ambiguity due to
gauge freedom.) If the two coupling constants are the same then Einstein
gravity is recovered. In this case a massless scalar is, if any, not a 
Brans-Dicke scalar but just a Klein-Gordon massless scalar. Note that 
none of the above equations (\ref{eqn:GN}) for zero modes says
anything about relations between metric and matter perturbations since
the left hand sides vanish. Hence, the zero mode does not affect the
value of $G_{N}^{(S,T)}$ directly. If the two coupling constants are
different and if there is a massless scalar then the weak gravity at low
energy is Brans-Dicke type. Finally, if the two coupling constants are
different and if there is no massless scalar then we do not have a
$4$-dimensional covariant description of the system: our intuition based
on $4$-dimensional gravity just breaks down. Therefore, in principle we
have four cases: (i) Einstein gravity without massless scalars; (ii)
Einstein gravity with massless scalars; (iii) Brans-Dicke gravity; (iv)
no $4$-dimensional covariant description.

In this paper, we shall review the result that (i) is the case for a 
model with one bulk scalar~\cite{Mukohyama-Kofman} and that high-energy
corrections to Einstein gravity can be mimicked by higher derivative
terms~\cite{Mukohyama}. Of course,  for the original Randall-Sundrum
model without bulk scalars, (iii) is the case~\cite{Garriga-Tanaka}. For
a particular class of perturbations, it has been shown that these
conclusions are true even for second order in
perturbations~\cite{Kudoh-Tanaka}. It is fortune that we do not
encounter the cases (ii) or (iv) for models without or with just one 
bulk scalar. However, for models with more than one bulk scalars, we do
not know whether we still have the same fortune or not.


\section{Formalism}

Throughout this paper we consider a $5$-dimensional brane world model
with a bulk scalar field $\Psi$ between two end-of-the-world branes 
$\Sigma_{\pm}$. The scalar field has potential in the bulk $U$ and on
the branes $V_{\pm}$. The positions of two branes are described by the
parametric equations
%
\begin{equation}
 x^M=Z^M_{\pm}(y_{\pm}),
\end{equation} 
where $x^M$ ($M=0,\cdots,4$) are $5$-dimensional coordinates in the
bulk, and $y_+$ and $y_-$ represent $4$-dimensional coordinate systems
on the branes $\Sigma_+$ and $\Sigma_-$, respectively. We can consider
coordinate transformations of $x^M$ and $y_{\pm}^{\mu}$
independently. In this sense, this setting is doubly covariant.

\begin{figure}[b]
\centering\leavevmode\epsfysize=8cm \epsfbox{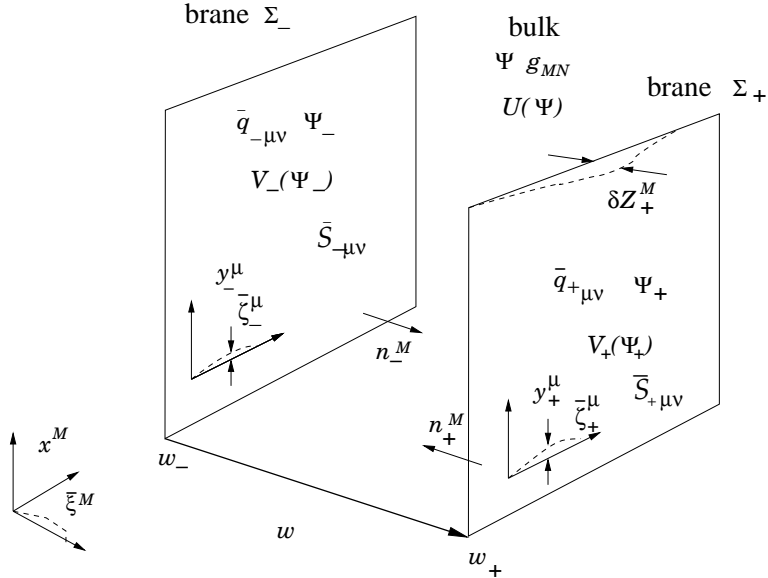}
\caption{\label{fig1} Sketch of the brane world geometry.}
\end{figure}

What we are interested in is classical dynamics of perturbations around
a background with $4$-dimensional Poincare symmetry since it determines
the effective $4$-dimensional theory of weak gravity on the
branes. Those perturbations include perturbations of $Z_{\pm}^M$ as well
as those of the metric $g_{MN}$, the scalar field $\Psi$, and matter on
the branes. We assume that there is no matter excitation on the brane 
$\Sigma_-$, considering it as the hidden brane and the other brane
$\Sigma_+$ as our brane. Equations which we can use are the
$5$-dimensional Einstein equation, the $5$-dimensional scalar field
equation, the Israel metric junction condition~\cite{Israel} and the
scalar field junction condition. The setting is summarized in
figure~\ref{fig1}. 

An important point is that the induced metric $q_{\pm\mu\nu}$ is not
necessarily minimally coupled to matter on the brane in the sense that
the matter action on the brane may depend on the pullback $\Psi_{\pm}$
of the scalar field. We assume that the minimally coupled metric
$\bar{q}_{\pm\mu\nu}$ is conformally related to the induced metric: 
%
\begin{equation}
 \bar{q}_{\pm\mu\nu}=e^{-\alpha_{\pm}(\Psi_{\pm})}q_{\pm\mu\nu}. 
\end{equation} 
In other words, we assume that the matter action depends on the pullback
of the scalar field via this conformal factor only. Correspondingly, the
physical stress energy tensor of matter on the brane should be defined
by 
%
\begin{equation}
 \bar{S}^{\mu\nu}_{\pm} = \frac{2}{\sqrt{-\bar{q}_{\pm}}}
  \frac{\delta}{\delta\bar{q}_{\pm\mu\nu}}I_{\pm matter}.
\end{equation} 
Note that this is different from the surface stress energy tensor which
is used in the Israel metric junction condition. There is a simple
relation between two stress energy tensors. 

Hereafter, we consider modes with non-vanishing $4$-momentum
$k_{\mu}$. Actually, a mode with vanishing $4$-momentum corresponds to
just a change of background within the $4$-dimensional Poincare
symmetry. On the other hand, we consider modes with vanishing
$k^{\mu}k_{\mu}$ as far as $k_{\mu}$ itself is not zero.

\begin{figure}[b]
\centering\leavevmode\epsfysize=6cm \epsfbox{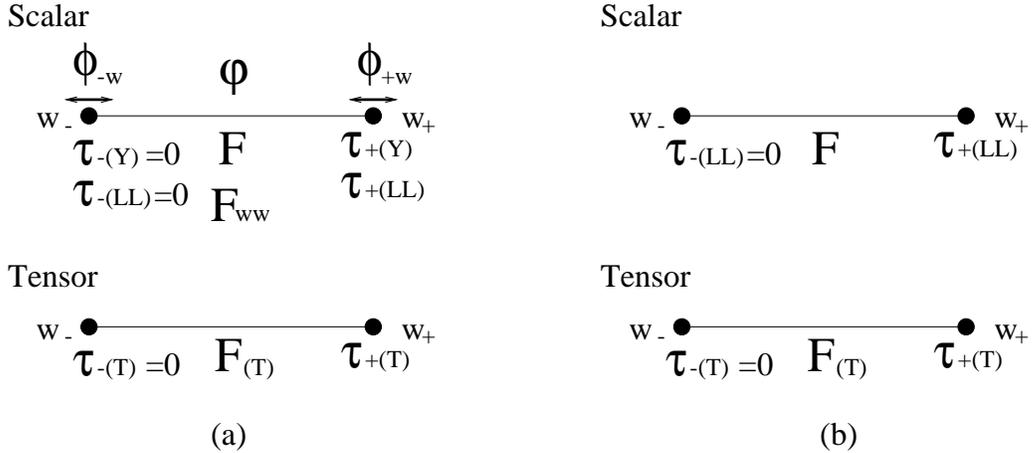}
\caption{\label{fig2} Simple $1$-dimensional problems.
}
\end{figure}

In order to analyse the system, let us (i) perform Fourier
transformation with respect to the $4$-dimensional coordinates; (ii)
classify Fourier coefficients into scalar, vector and tensor parts (each
part forms an irreducible representation of the little group of 
$4$-dimensional Poincare symmetry and, thus, can be analysed
independently); (iii) construct gauge-invariant variables in each
part. After these three steps, we have a set of purely $1$-dimensional
problems as shown in figure~\ref{fig2}~(a) and in the following three
paragraphs. 

For the scalar part, we have three gauge invariant variables in the bulk
and three gauge invariant variables on each brane. The variables $F$ and
$F_{ww}$ are constructed from bulk metric perturbations only, while
$\varphi$ is constructed from perturbations of the scalar field and
the metric. The variable $\phi_{\pm w}$ is a gauge invariant variable
corresponding to the normal component of perturbations of the brane
position. Other gauge invariant variables $\bar{\tau}_{\pm(Y)}$ and
$\bar{\tau}_{\pm(LL)}$ correspond to perturbations of the physical
stress energy tensor of matter on the brane. Since it was assumed that
there is no excitation of matter on the hidden brane,
$\bar{\tau}_{-(Y)}$ and $\bar{\tau}_{-(LL)}$ are zero. For the three
variables in the bulk we have one differential equation and two
algebraic equations. Hence we can eliminate, for example, $\varphi$ and
$F_{ww}$. As for the three variables on the brane, the metric junction
condition gives a conservation equation and an expression of $\phi_{\pm
w}$ in terms of $\bar{\tau}_{\pm(LL)}$. The conservation equation is a
linear relation between $\bar{\tau}_{\pm(Y)}$ and
$\bar{\tau}_{\pm(LL)}$. Hence, we can eliminate $\phi_{\pm w}$ and
$\bar{\tau}_{\pm(Y)}$. Finally, the scalar field junction condition
gives a boundary condition of bulk perturbations at $w=w_{\pm}$
depending on $\bar{\tau}_{+(LL)}$.

The vector part is actually irrelevant since vector perturbations vanish
if matter perturbations on the hidden brane vanish.

As for the tensor part, we have one differential equation for one
variable in the bulk, and the metric junction condition gives a boundary
condition at $w=w_{\pm}$ depending on $\bar{\tau}_{+(T)}$.

Therefore, we finally simplified the $5$-dimensional problem to a set of
simple $1$-dimensional problems shown in figure~\ref{fig2}~(b). For each
sector of scalar and tensor parts, we have one differential equation
with a boundary condition depending on matter on our brane. This system
may be simple enough to analyse. If we can solve the $1$-dimensional
differential equations with the boundary conditions then we can
calculate gauge-invariant perturbations $\bar{f}$ and $\bar{f}_{(T)}$ of
the $4$-dimensional physical metric on our brane. Since the boundary
condition depends linearly on matter on our brane, the result should be
of the form
%
\begin{eqnarray}
 C_{(S)}\bar{f}_+ & = & \bar{\tau}_{+(LL)}, \nonumber\\ 
 \bar{q}_+^{(0)\mu\nu}k_{\mu}k_{\nu}C_{(T)}\bar{f}_{+(T)} 
  & = & \bar{\tau}_{+(T)},
  \label{eqn:expectation}
\end{eqnarray}
where $C_{(S)}$ and $C_{(T)}$ are functions of
$\bar{q}_+^{(0)\mu\nu}k_{\mu}k_{\nu}$ and 
$\bar{q}_+^{(0)\mu\nu}=\Omega_+^{-2}\eta^{\mu\nu}$ is the inverse of the
unperturbed physical metric $\bar{q}_{+\mu\nu}^{(0)}$. The reason why 
$\bar{q}_+^{(0)\mu\nu}k_{\mu}k_{\nu}$ was put in front of
$\bar{f}_{+(T)}$ is that we expect the emergence of $4$-dimensional 
gravitons on our brane (non-vanishing $\bar{f}_{+(T)}$ with
$\bar{q}_+^{(0)\mu\nu}k_{\mu}k_{\nu}=0$ and $\bar{\tau}_{+(T)}=0$). This
is actually the case~\cite{Mukohyama-Kofman}.

What is important here is that the functions $C_{(S,T)}$ completely 
characterize the effective theory of weak gravity on our
brane. Actually, if one likes, one can restore gauge fixed equations for
any gauge choices from the functions $C_{(S,T)}$ only. Since we are
dealing with gauge-invariant variables only, there is no ambiguity of
gauge freedom when we compare the result with the corresponding
equations in a $4$-dimensional gravitational theory. Namely, we only
have to compare functions $C_{(S,T)}$ of 
$\bar{q}_+^{(0)\mu\nu}k_{\mu}k_{\nu}$ with the corresponding functions
of momentum squared in the Fourier transformed, linearized
$4$-dimensional gravitational theory.

In order to solve the reduced $1$-dimensional problems, we expand them
by the dimensionless parameter $\mu=l^2\eta^{\mu\nu}k_{\mu}k_{\nu}$ so
that we can solve equations iteratively, where $l$ is the characteristic
length scale of the model which we can determine by comparing the
results of order $O(1)$ and $O(\mu)$. The purpose of the $\mu$-expansion
is to analyze the behavior of the functions $C_{(S)}$ and $C_{(T)}$ near 
$\mu=0$. Namely, we shall seek first few coefficients $C_{(S,T)}^{[i]}$
($i=0,1,\cdots$) of the expansion 
%
\begin{equation}
 C_{(S,T)} = \sum_{i=0}^{\infty}\mu^iC_{(S,T)}^{[i]}. 
  \label{eqn:expansion-C}
\end{equation}
Since the $4$-dimensional physical energy scale $m_+$ on $\Sigma_+$ is
given by $m_+^2$ $=$ $-\bar{q}_+^{(0)\mu\nu}k_{\mu}k_{\nu}$ 
$\propto$ $-\mu l^{-2}$, the expansion in $\mu$ is nothing but the low
energy expansion. Hence, the first few coefficients $C_{(S,T)}^{[i]}$
($i=0,1,\cdots$) of the expansion (\ref{eqn:expansion-C}) determine the
low energy behavior of weak gravity on our brane.


\section{Result and conjecture}

In the low energy expansion (\ref{eqn:expansion-C}), we expect that
$C_{(S,T)}^{[0]}$ give $4$-dimensional Einstein gravity, that
$C_{(S,T)}^{[1]}$ give curvature-squared corrections to the Einstein
gravity, and so on. Actually, this is the case. The effective Newton's
constant was determined in ref.~\cite{Mukohyama-Kofman} and coefficients
of curvature-squared terms were determined in ref.~\cite{Mukohyama}. For
a trivial $4$-dimensional conformal transformation between the induced 
metric and the physical metric ($\alpha_{\pm}=0$), the Newton's constant
agrees with the result of ref.~\cite{Tanaka-Montes}. In
ref.~\cite{Tanaka-Montes}, the freedom of the conformal transformation 
was not considered. On the other hand, in
refs.~\cite{Mukohyama-Kofman,Mukohyama} a conformal transformation
depending arbitrarily on the bulk scalar field was taken into account.

Equipped with the result up to the order $O(\mu^1)$, we conjecture
that in the order $O(\mu^N)$, weak gravity on the brane should still be 
indistinguishable from a higher derivative gravity whose action includes
up to $2(N+1)$-th derivatives of metric. Two linear combinations of
coefficients of higher derivative terms can be fixed by using the
iterative results in ref.~\cite{Mukohyama}. Since the expansion in $\mu$
is in principle an infinite series, gravity in the brane world becomes
non-local at high energies even at the linearized level. Actually, in
this case equations of brane gravity will have up to $2(N+1)$-th
derivatives ($N\to\infty$) and, thus, we need to specify the $0$-th to
$(2N+1)$-th time-derivatives ($N\to\infty$) of the metric perturbations
on a spacelike $3$-surface on the brane in order to predict the future
evolution. In other words, we need to specify the whole (past) history
of perturbations on the brane to predict the future evolution with 
infinitely high accuracy, provided that the metric perturbations can be
Taylor expanded with respect to the time. This explains how the
$4$-dimensional local description breaks down at high energies. Of 
course, at low energies the non-local behavior is well suppressed. The
result and the conjecture are summarized in table~\ref{table1}. 
%
\begin{table}[hbt]
\caption{Result and conjecture}
        \label{table1}
\begin{center}
\begin{tabular}{|c|c|c|} \hline
 & $4D$ effective theory \\ \hline 
 $O(\mu^0)$ & Einstein theory\\ \hline
 $O(\mu^1)$ & $R^2$-theory\\ \hline
 \vdots & \vdots \\ \hline
 $O(\mu^N)$ & $R\Box^{N-1}R$-theory\\ \hline
 \vdots & \vdots \\ \hline
 $O(\mu^{\infty})$ & Nonlocal theory\\ \hline
\end{tabular}
\end{center}
\end{table}

%
\begin{figure}[hbt]
 \caption{Effects of bulk gravitational/scalar waves} 
 \label{fig:bulk-wave}
 \setlength{\unitlength}{1mm}
 \begin{center}
  \begin{picture}(103,34)
   \put(13,27){\framebox(46,7){%
   Perturbations on the brane}}
   \put(70,17){\framebox(35,10){%
   \shortstack{Gravitational/scalar\\
   waves in the bulk}}}
   \put(14,7){\framebox(43,7){%
   Interactions (collision)}}
   \put(63,28){generate}
   \put(63,13){propagate in the bulk}
   \put(37,18){%
   \shortstack{propagate\\
   on the brane}}
   \put(37,2){continue}
   \put(0,28){
   \shortstack{%
   $t=t_1$\\
   ${\bf x}={\bf x}_1$}}
   \put(0,8){
   \shortstack{%
   $t=t_2$\\
   ${\bf x}={\bf x}_2$}}
   \put(59,30){\vector(3,-2){11}}
   \put(70,21){\vector(-3,-2){13}}
   \put(32,27){\vector(0,-1){13}}
   \put(32,7){\vector(0,-1){7}}
  \end{picture}
  \end{center}
\end{figure}
The non-locality is due to gravitational and scalar waves in the bulk as
shown in figure~\ref{fig:bulk-wave}. First, let us consider
perturbations localized on the brane. By definition, these perturbations
propagate on the brane. At the same time, they can generate
gravitational and scalar waves since the junction conditions couple
perturbations on the brane and those in the bulk. These waves, of
course, propagate in the bulk spacetime. However, they may eventually
collide with the brane, and can alter the evolution of perturbations
localized on the brane. Hence, from the $4$-dimensional point of view,
the evolution of perturbations localized on the brane should be
non-local. The non-locality is one of essential features of brane
worlds.

Hence, the infinite series of higher derivative terms is one description
of the non-locality pointed out in ref.~\cite{Mukohyama2000b} in the
context of brane world cosmology~\cite{Cosmology}. Another description
was given as an integro-differential equation for the brane world
scenario without bulk scalars in ref.~\cite{Mukohyama2001a}, where a
complete set of four equations governing scalar-type cosmological
perturbations was derived by using the doubly gauge invariant 
formalism~\cite{Mukohyama2000b,Mukohyama2000c,Mukohyama-jgrg}. One of
the four equations is an integro-differential equation, which describes 
non-local effects due to gravitational waves propagating in the
bulk. Further investigation of the relation between the two different
descriptions may be an interesting future subject. For example, the
infinite series of higher derivative terms in the original
Randall-Sundrum model with the infinite fifth dimension without bulk
scalars was obtained in ref.~\cite{GKR}. It should also be possible to
obtain integro-differential equations for the model with a bulk scalar.

\section*{Acknowledgements}
 Part of contents presented here is based on collaboration with Lev
 Kofman. The author would like to thank him for useful discussions and
 comments. This work is supported by JSPS Postdoctoral Fellowship for
 Research Abroad.

\end{document}